\documentclass[a4paper,fleqn,usenatbib]{mnras}

\usepackage{newtxtext,newtxmath}

\usepackage[T1]{fontenc}
\usepackage{ae,aecompl}

\usepackage{mathtools}
\usepackage{xcolor, soul}
\sethlcolor{yellow}

\usepackage{graphicx}	
\usepackage{amsmath}	

\newcommand{\tone}{\theta_1}
\newcommand{\ttwo}{\theta_2}
\newcommand{\zd}{z_d}
\newcommand{\zs}{z_s}
\newcommand{\zds}{z_{ds}}
\newcommand{\dd}{D_d}
\newcommand{\ds}{D_s}
\newcommand{\ddss}{D_{ds}}
\newcommand{\dds}{D_{ds}{ \left(z_d, z_s, \Omega_{\Lambda, 0} \right) }}
\newcommand{\dt}{\Delta\tau}
\newcommand{\df}{D_f}
\newcommand{\Hub}{H_0}
\newcommand{\veld}{\sigma_v}
\newcommand{\veldc}{\frac{\veld}{c}}
\newcommand{\veldcs}{\left(\veldc\right)^2}
\newcommand{\veldpi}{\frac{{4}\pi}{c}}
\newcommand{\veldcsc}{{4}\pi\veldcs}
\newcommand{\veldcscs}{\left[\veldcsc\right]^2}
\newcommand{\ddd}{\frac{\dd\ddss}{\ds}}
\newcommand{\etan}{\eta_0}
\newcommand{\xin}{\xi_0}
\newcommand{\Hubc}{\frac{c}{\Hub}}
\newcommand{\zob}{z^{observed}}
\newcommand{\fo}{\frac{1}{3}}
\newcommand{\ft}{\frac{1}{2}}
\newcommand{\ftt}{\frac{4}{3}}
\newcommand{\fts}{\frac{3}{8}}
\newcommand{\lts}{\frac{{1+\zd}}{{1+\zs}}}
\newcommand{\onez}{\frac{1}{{1+z}}}
\newcommand{\onezs}{\frac{1}{{1+\zs}}}
\newcommand{\mat}{\Omega_{\textit{m},0}}
\newcommand{\rad}{\Omega_{\textit{r},0}}
\newcommand{\dark}{\Omega_{{\Lambda},0}}
\newcommand{\any}{\Omega_{\textit{i},0}}
\newcommand{\Psis}{\Psi\left(\zs, \dark\right)}
\newcommand{\Psid}{\Psi\left(\zd, \dark\right)}
\newcommand{\Psin}{\Psi\left(z, \dark\right)}
\newcommand{\darks}{\sqrt{\dark}}
\newcommand{\bc}{\beta\cos\epsilon}
\newcommand{\Hypt}{_2F_1}
\newcommand{\darko}{\frac{1}{\dark}}
\newcommand{\darkos}{\frac{1}{\darks}}
\newcommand{\Intone}{\int\limits_{\onez}^{1}\frac{dx}{\sqrt{x^4 \, \dark + x\, \mat + \rad }}}
\newcommand{\Inttwo}{\int\limits_{\lts}^{1}\frac{dx}{\sqrt{x^4 + x\left(\frac{1}{\dark}-1\right)(1+\zd)^3}}}

\newcommand{\Delay}{\frac{\df}{c}(1+\zd)\left[\; \frac{1}{2}(\tone^2-\ttwo^2)+|\tone\ttwo| \; \ln\left|\frac{\tone}{\ttwo}
	\; \right|\;\right]}
\newcommand{\red}{\frac{\sqrt{1-\beta^2}}{1-\bc}(1+\zob)}
\newcommand{\sisd}{{c}{\dt}}

\title[Gravitational Lensing Time Delay with Peculiar Motions]{The Effect of the Peculiar Motions of the Lens, Source and the Observer on the Gravitational Lensing Time Delay}

\author[Weerasekara et al.]{
Gihan Weerasekara,$^{1}$\thanks{E-mail: contactgihan@gmail.com}
Thulsi Wickramasinghe,$^{2}$
Chandana  Jayaratne$^{1}$
\\
$^{1}$Department of Physics, University of Colombo, Sri Lanka\\
$^{2}$Department of Physics, The College of New Jersey, Ewing, NJ 08628, USA\\
}

\date{Accepted XXX. Received YYY; in original form ZZZ}

\pubyear{2018}

\begin{document}
\label{firstpage}
\pagerange{\pageref{firstpage}--\pageref{lastpage}}
\maketitle

\begin{abstract}
An intervening galaxy acts as a gravitational lens and produces multiple images of a single source such as a remote galaxy. 
Galaxies have peculiar speeds in addition to the bulk motion arising
 due to the expansion of the universe. 
 There is a difference in light arrival times 
 between lensed images. 
 We calculate more realistic time delays
 between lensed images when galaxy
 peculiar motions, that is the motion of the Lens, the Source and the Observer are taken into consideration neglecting the gravitomagnetic effects.
\end{abstract}




\section{Introduction}

A remote galaxy S at redshift $z_s$ 
(Shown in Figure 1) is lensed by an intervening galaxy L at redshift $\zd$. 
A light ray from S  bends by an angle $\alpha$ before arriving at the observer O. The image I of S forms at an angle $\theta$ while S is at $\beta$.
 The distances  $\dd$,  $\ds$ and $D_{ds}$ shown are the angular diameter distances. \citet{Walsh1979}, \citet{Chen1995}

From the theory of lensing,
 we can derive the angular 
 positions  $\theta_1$ and $\theta_2$
  of the 
   two  lensed images formed due to  a single 
   \emph{point} 
   lens.
   There is a  delay $\dt$ of light arrival times from these two 
   images. 
   This delay is arising due to both geometrical path difference and 
   the fact that two light rays are traveling in two different potential wells on either side of the lens. 
   The total time delay is given by, \citet{Schneider1992}, \citet{Bradt2008}

\begin{equation}
\dt =  \Delay
\end{equation}
where,

\begin{equation}
D_f = \frac{ \dd   \ds }{D_{ds} } 
\end{equation}
We calculate analytically a more realistic time delay between the two images when  the peculiar speeds of the lens, the source and the observer are considered. These peculiar speeds are random speeds With respect to the cosmic microwave background radiation - Hubble flow.

But as we already know a point mass lens is a highly idealized and less practical lensing model for a real lensing system, in the next part of the paper we will be considering a more practical Singular Isothermal Sphere (SIS) lensing model to calculate the time delay difference when the peculiar speeds of the objects are considered.

\begin{figure}
	\includegraphics[width=\columnwidth]{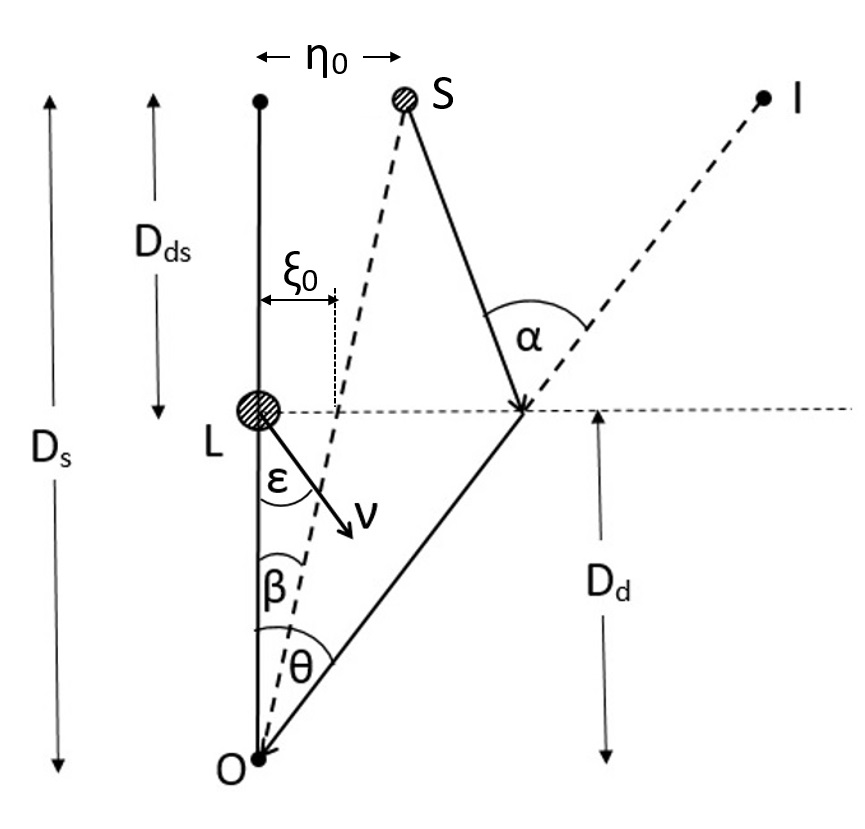}
    \caption{Gravitational Lensing Diagram. The peculiar speed $v$ of the lens L is measured with respect to a freely falling 
observer with the Hubble flow at the location of the lens. 
The angle $\epsilon$ is measured from the optic axis OL. }
    \label{fig:example_figure}
\end{figure}

\section{Theory}

The angular diameter distance D of a source \textit{having no peculiar motion} at a red shift $z$  is given by, \citet{Weinberg1972}, \citet{Hobson2006}


\begin{equation}
D(z, \dark) = \Hubc \onez \Intone
\end{equation}
where $\any$ is the density parameter of the substance 
$i$ of the cosmic fluid 
 measured at the present time $t_0$. 
 We assume a flat universe 
($k=0$) for which \citet{Perlmutter1999}, 

\begin{equation}
\mat + \rad + \dark = 1
\end{equation}
\\
The red shift  $\zds$ of S as measured  by L is given by,

\begin{equation}
1+\zs = (1+\zd)(1+\zds)
\end{equation}
\\
Thus, from the equations (3), (4) and (5), neglecting $\rad$ and eliminating $\mat$ and expressing everything with the dark energy, we can derive  the value of $D_{ds}$, the angular diameter distance of the source as measured 
 by an observer on the lens as,


\begin{equation}
\begin{multlined}
\dds  = \\ \Hubc \darkos \lts \Inttwo
\end{multlined}
\end{equation}

By evaluating the integral analytically, 
the value of $D_{ds}$ can be written as

\begin{equation}
\dds = \Hubc \onezs
\left[ \Psis - \Psid \right]
\end{equation}
\\
where in terms of   hypergeometric function $_2F_1$ 

\begin{equation}
\Psin = \frac{1 + z}{\darks}\; 
	\Hypt\left(\fo, \ft; \ftt ;
	\left(1 - \darko \right)(1 + z)^3  
	 \right) 
\end{equation}
\\
In the theory of lensing, the 
source S, lens L,  and the observer O in Fig. 1 are all freely falling
with the smooth expansion of the universe;  that is, \textit{experiencing no peculiar motions}. The angular diameter distances $\ds$, $\dd$ and $D_{ds}$ are then  measured between these objects which are freely falling with the Hubble flow. Thus, the redshifts entering Eq (8) should be associated with the freely falling objects.

However, all galaxies are subjected to peculiar or random motions, for an example in the scenario given here the Source S, the Lens L and the Observer O are having peculiar motions. Thus, the redshift of the lens we measure includes this peculiar motion. Therefore, the redshifts entering Eq (7), which should be the redshifts of freely falling objects,  must be corrected for random peculiar motions. For this, consider initially the random motion of L neglecting the random motions of S and O. This is similar to OS axis being fixed and L having a peculiar motion with respect to this axis. An observer freely falling with the Hubble flow at the location of L will see a Doppler shift of L arising due to the random (peculiar) speed $\nu$. In addition to this shift, we have the cosmological redshift of that freely falling observer arising due to the bulk expanding motion of the universe. Thus, the redshift z of the freely falling observer, from special theory of relativity, becomes (see Figure. 1)
 
 \begin{equation}
1+z = \red
\end{equation}
\\
where $v = \beta c$ is the peculiar speed of the object
as seen by the freely falling observer and $\epsilon$ is the angle between the peculiar velocity vector and the line-of-sight to L (see Fig. 1). 
It is this redshift $z$ (Eq. 9) that should enter in (7) for the 
angular diameter distance calculation. 
If $\epsilon = 0$, L is approaching a freely falling 
observer and if $\epsilon = \pi$ it is receding.
Inserting (9) in (8) and expanding to first order in $\beta$ we get,

\begin{equation}
\begin{multlined}
\Psin \sim 
\frac{1 + \zob}{\darks} \; \times \\ \Hypt 
\left[ 1 +  \left\{ 1 + \fts  \left( 1 - \darko \right) 
\left(1 + \zob\right)^3 
 \right\}  \bc \; \right] 
\end{multlined}
\end{equation}
\\
where the hypergeometric function is the one 
appearing in (8) with $z = z^{observed}$. 
Now that we have an expression to account for the peculiar motion of L, we can employ the same in our code to calculate the time delay taking all the peculiar motions into consideration. That is including the peculiar motions of S, L and O.
while doing so, we find that the other higher order terms are very small and 
the time delay is \emph{linear} to first order in $\beta$. 
Then the form of the observed time delay becomes,

\begin{equation}
\dt \approx \dt_0\,  ( 1  + \kappa\;  \bc )
\end{equation}
\\
where $\dt_0$ is when the peculiar motions are neglected. 

As we now have an equation for the gravitational time delay difference when the peculiar speeds are considered for a point mass lens model, let us now proceed to the Singular Isothermal Sphere lensing model and derive the time delay difference equation for that.

According to the theory of lensing the time delay difference for a SIS model is given by the equation, \citet{Schneider1992} 

\begin{equation}
\sisd = \veldcscs \ddd (1+\zd) 2y
\end{equation}
further by making use of the following equations,

\begin{equation}
y = \frac{\eta}{\etan}
\end{equation}

\begin{equation}
\xin = \veldcsc \ddd
\end{equation}
we can arrive at the following equation that gives us the required time delay.

\begin{equation}
\dt = \veldpi \veldcs \dd (1+\zd) 2\beta
\end{equation}
we do a realistic assumption for $\beta$ by making use of the point mass lens model as,

\begin{equation}
\beta = \tone + \ttwo
\end{equation}

{In this equation when we consider the peculiar speeds of the objects, we have to use $z = z^{observed}$ in accordance with (9) similar to the calculation we have carried out with the point mass lens. 

\section{Results and Discussion}

The example we have used is the lensing system illustrated in the Figure 2. \citet{Koopmans1998} This lens is referred to as B1600+434 and it has the following characteristics.
\vspace{0.2cm}

Optical time delay \hspace{1cm}= 51 $\pm$ 2 Days

\vspace{0.2cm}

$\zs$ \hspace{0.5cm}= 1.59 \hspace{2cm}$\tone$ \hspace{0.5cm}= +1.14"

$\zd$ \hspace{0.46cm}= 0.42 \hspace{2cm}$\ttwo$ \hspace{0.5cm}= -0.25"
\vspace{0.2cm}

\begin{figure}
	\includegraphics[width=\columnwidth]{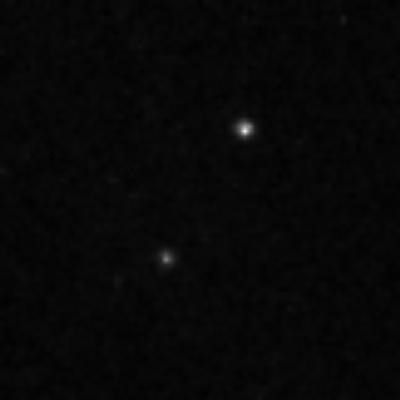}
    \caption{Lensing image. The Optical and Radio delay for this system has been measured. \citet{Koopmans1998} }
    \label{fig:example_figure}
\end{figure}

According to the given set of angular distances and angles assuming the 
\emph{non-realistic} 
assumption  that the lens is a point mass, we can calculate a theoretical lensing delay time of 73.92 days for the WMAP  cosmological parameters.
When we compare the theoretical time delay and the observed time delays it is clear that they are not matching.
We believe that the discrepancy is arising due to the lens point-mass assumption 
and  that we have not taken peculiar speeds into account. 
 However we would like to illustrate the effect of the peculiar motions on the time delay assuming 
 initially a point-mass lens here.

We simulated 
 1000 scenarios with the above given particular set of lensing parameters ($\zs$ = 1.59, $\zd$ = 0.42, $\tone$ = +1.14" and $\ttwo$ = -0.25" ). 
 For each scenario the lens and the observer have 
  random peculiar speeds in random directions with respect to the back ground radiation. In the simulations of Figure 3/4/5. the peciliar speeds are non relativistic and they range from 0 to 0.01$c$.

  for this lensing system Eq (11) can be written as,
  
\begin{equation}
\dt \approx 73.92 \,  ( 1  + 4.69\;  \bc )
\end{equation}

The observer, that is the Milky Way has an estimated peculiar speed of $600kms^{-1}$ \citet{Kogut1993} with respect to the back ground radiation. The directions of the peculiar motions are taken to be random in relation to the OL axis. We have taken $\dark$ = 0.73.

The simulated time delays as shown in Figure 3. are showing a time delay range of 8 days with the contribution of the peculiar motions while no peculiar motion time delay being 73.9 days. Therefore the maximum time delay when all three objects are moving is nearly 4 days and it is a significant value. Therefore the peculiar motions will give rise to a measurable and significant difference in the gravitational lensing time delay.

\begin{figure}
	\includegraphics[width=\columnwidth]{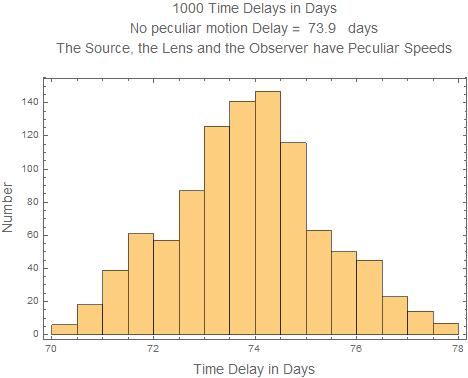}
    \caption{Point Mass lens. The Source, the Lens, and the Observer all are having peculiar speeds in the range of 0 to 0.01c in any random direction}
    \label{fig:example_figure}
\end{figure}

In the second simulation given in Figure 4 we have excluded only the peculiar motion of the Lens. In this case it is seen that the maximum time delay difference is about 1 day. From this result it is clear that the 
peculiar motions of the source and the Observer alone when the lens is not moving is not creating a significant gravitational lensing time delay. To further enhance this fact we have taken another simulation with only the Lens having peculiar motions and the observer and the source are stationary. That result is given in the Figure 5.

\begin{figure}
	\includegraphics[width=\columnwidth]{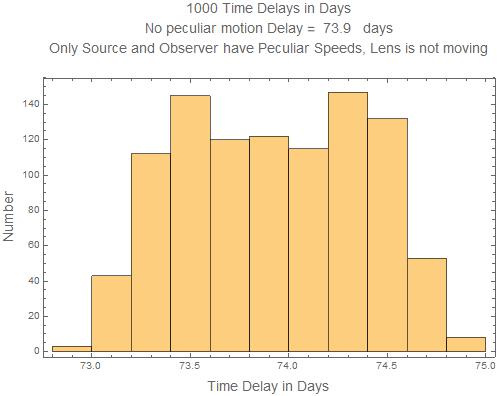}
    \caption{Point Mass lens. The Source and the Observer are having peculiar speeds in the range of 0 to 0.01c in any random direction. The Lens is stationary }
    \label{fig:example_figure}
\end{figure}

\begin{figure}
	\includegraphics[width=\columnwidth]{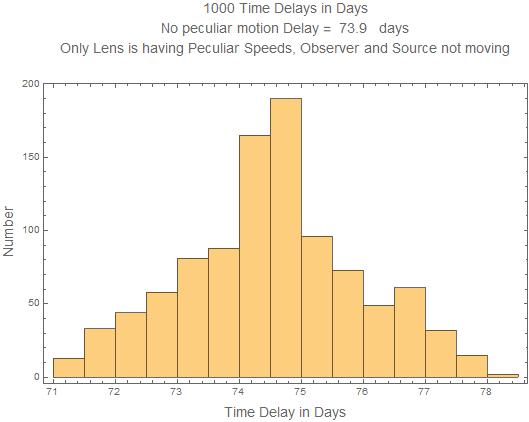}
    \caption{Point Mass lens. The Lens is having peculiar speeds in the range of 0 to 0.01c in any random direction. The source and the observer are stationary }
    \label{fig:example_figure}
\end{figure}

The result we have obtained in Figure 5 is almost identical to the result we have obtained in the Figure 3.  

From these results it is clear that the gravitational lensing time delay is highly sensitive to the peculiar speeds of the lens. Another interesting result of the simulation is the peculiar speeds of the observer and the source is not having a significant effect on the gravitational lensing time delay.

As we have figured out by now, the gravitational lensing time delay is mostly affected by the peculiar motions of the Lens. Thus we can neglect the peculiar motions of the Observer and the Source.

In the next simulation given in Figure 6, we have taken a lensing system with only the lens moving. In that we have taken the speed and the direction of the lens separately. The lens in the simulation is having speeds from 0 to 0.005$c$ and the direction is
 0  (The lens is approaching the observer) to $\pi$ 
  (The lens is receding from the observer). 
  If the $\epsilon$
   is $\pi$/2 then the Lens is moving in a transverse direction.

From Figure 6, we can identify that when the lens is moving towards the observer the gravitational lensing time delay is increasing and it is attaining larger values directly in proportion with the peculiar speed of the lens. That is, when the lens is having larger approaching peculiar speeds the gravitational lensing time delay is also larger.

In contrast to that when the lens is receding from the observer the gravitation lensing time delay is decreasing. It can be also seen that when the receding peculiar speed is becoming larger the gravitational lensing time delay is becoming smaller.

If the lens is moving in a transverse direction then there is no measurable effect in the gravitational lensing time delay as the effect is in second order.

\begin{figure}
	\includegraphics[width=\columnwidth]{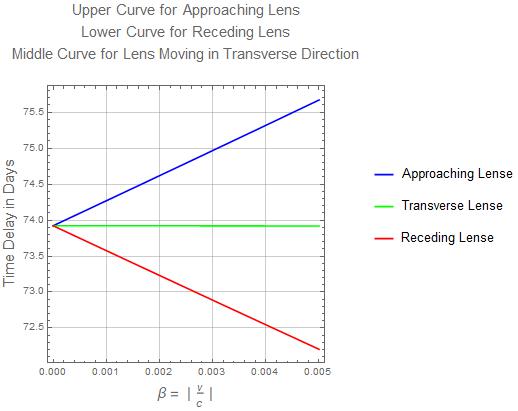}
    \caption{Point Mass lens. The lens is having different peculiar speeds in different directions }
    \label{fig:example_figure}
\end{figure}

The lenses we have considered so far are having small velocities. But if we consider lenses having relativistic speeds then the effect become more prominent. That is the measurable gravitational lensing time delay becomes much larger. Results are illustrated in the Figure 7, where the peculiar speeds of the lens are relativistic.

\begin{figure}
	\includegraphics[width=\columnwidth]{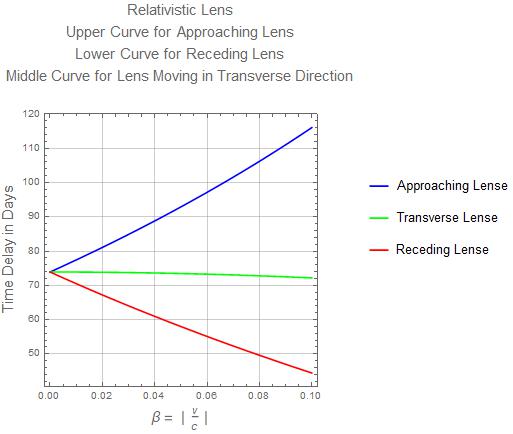}
    \caption{Point Mass lens. The Lens is having relativistic peculiar speeds }
    \label{fig:example_figure}
\end{figure}

In the example we have taken, the Lens B1400+434 is having an measured optical time delay of 51 days and a theoretical time delay of 73.92 days, assuming a point-mass lens. 
 From our results we can account for the difference of this time delay. That is we can have this particular observed optical time delay difference if the lens is having a relativistic peculiar speed in the range of 0.05$c$ to 0.06$c$ in a receding direction from us provided that we model the lens as a point mass, which is \emph{ not exact.}
 
As we now have a clear idea on gravitational lensing time delays when the peculiar speeds of the objects are considered while using a point mass lensing model, let us now investigate the same effect when a more realistic Singular Isothermal Sphere lensing model is used for the calculations.

For this also we employ the same simulation with 1000 scenarios where random peculiar speeds are in random directions. when using Eq. (15) average velocity dispersion $\veld$ will be taken as $150 kms^{-1}$ \citet{Koopmans1998}. With this average velocity dispersion value and using Singular Isothermal Sphere model we have a very interesting result for the non peculiar motion lensing time delay, which is 51.45 days. this value is almost identical to the observed lensing time delay value of 51 $\pm$ 2 Days.

The simulation for the non relativistic peculiar speeds is given in the Figure 8. In that the non relativistic peculiar speeds are from 0 to 0.01$c$. further it can be noted in this simulation the time delays are ranging from 50.5 - 52.5 days while having a maximum delay difference of 1 day from the no peculiar motion instance. therefore even with non relativistic peculiar speeds it is clear that we can have measurable and significant time delay difference from the no peculiar motion instance when peculiar speeds of the lens is considered.

In the next simulation given in the Figure 9. we consider a relativistic peculiar speed distribution from 0 to 0.05$c$. it can be noted in this figure when there is a relativistic peculiar speed distribution for the lens, the lensing time delays can range from 46-56 days with a maximum delay difference of 5 days from the no peculiar motion instance.
therefore it is apparent from this simulation when there is a relativistic peculiar speed for the lens there can be a very significant gravitational lensing time difference from the non peculiar speed instance while using a more realistic Singular Isothermal Sphere to model the lens.

\begin{figure}
	\includegraphics[width=\columnwidth]{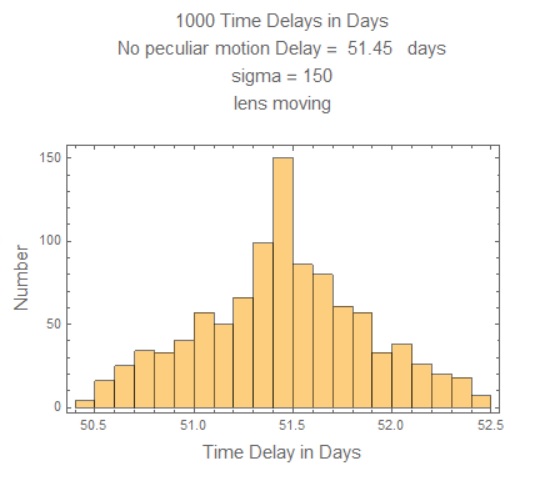}
    \caption{Singular Isothermal Sphere lens model. The Lens is having non relativistic peculiar speeds in the range of 0 to 0.01c in any random direction }
    \label{fig:example_figure}
\end{figure}

\begin{figure}
	\includegraphics[width=\columnwidth]{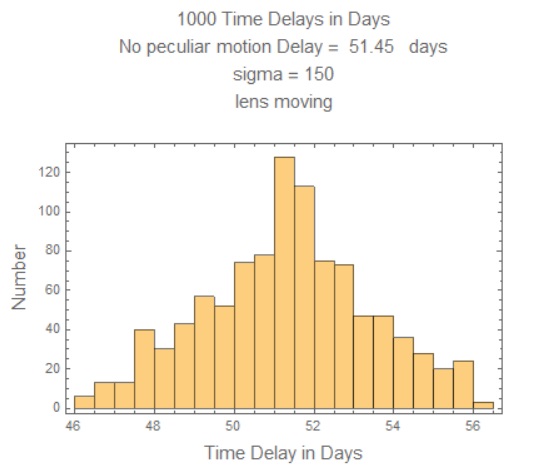}
    \caption{Singular Isothermal Sphere lens model. The Lens is having relativistic peculiar speeds in the range of 0 to 0.05c in any random direction }
    \label{fig:example_figure}
\end{figure}










\section{Conclusions}

From the above simulations we have found out that in fact there is a significant measurable time delay difference arising from the peculiar speeds of the lens using both non realistic point mass lens and more realistic Singular Isothermal Sphere as the lensing model.

The important observation is that an approaching lens results in an increase of the time delay 
while a receding lens gives rise to a decrease in the delay.

We find that the time delay is not significantly affected by the source or observer peculiar motions.

We see from Figure 7. and Figure 9. that a relativistically moving lens in any direction can significantly affect 
the lensing time delays.

\section*{Data availability}

The data underlying this article will be shared on reasonable request
to the corresponding author.














\bsp	
\label{lastpage}
\end{document}